\documentclass[12pt]{iopart}
\usepackage{epsfig}
\usepackage{graphicx}
\usepackage{amssymb}

\begin{document}

\title{Nonextensive statistical effects and strangeness production in hot and dense nuclear matter}

\author{A Lavagno and D Pigato}
\address{Department of Applied Science and Technology, Politecnico di Torino, Italy and \\
Istituto Nazionale di Fisica Nucleare (INFN), Sezione di Torino, Italy}

\begin {abstract}
By means of an effective relativistic nuclear equation of state in the framework of the nonextensive statistical mechanics, characterized by power-law quantum distributions, we study the phase transition from hadronic matter to quark-gluon plasma at finite temperature and baryon density.
The analysis is performed by requiring the Gibbs conditions on the global conservation of baryon number, electric charge fraction and zero net strangeness.
We show that nonextensive statistical effects strongly influence the strangeness production during the pure hadronic phase and the hadron-quark-gluon mixed phase transition, also for small deviations from the standard Boltzmann-Gibbs statistics.
\end {abstract}

\section{Introduction}\label{intro}

One of the major challenges in high energy heavy ion collisions is the determination of the physical
proprieties of strongly interacting nuclear matter at high temperature and baryon density.
Relativistic heavy-ion collisions provide the unique possibility to explore in laboratory nuclear matter under extreme regimes in which the baryon density can reach values of a few times the saturation nuclear density and/or high temperatures. In these conditions, phase transition phenomena in the hot and dense fireball created during the collisions can take place \cite{hwa}.
Various results from QCD inspired models indicate that, increasing the baryon chemical potential in
the phase diagram, a region of non-singular but rapid cross-over of thermodynamic observable around a quasi-critical temperature, leads to a critical endpoint (CEP), beyond which the system shows
a first order phase transition from confined to deconfined matter.
The existence or exclusion of a CEP has not yet been confirmed by QCD lattice simulations. Actually, there are some extrapolation techniques to finite chemical potentials, although
the precise location of the CEP is still a matter of debate \cite{schaefer,fodor}. Such a CEP can be in principle detected in future high-energy compressed nuclear matter experiments such as FAIR at GSI in Darmstadt \cite{senger} and NICA at JINR in Dubna \cite{nica}. In this direction interesting results have been obtained at low SPS energy and are foreseen at a low-energy scan at RHIC \cite{alt,agga,richa}.

The process of deconfinement and the equation of state (EOS) of hot and dense nuclear matter can in principle be described by QCD, however such a theory is highly non-perturbative in the energy density range involved in relativistic heavy-ion collisions.
The generated quark-gluon plasma (QGP) in the early stages of the collisions does not at all resemble a quasi-ideal gas of quarks and gluons because strongly dynamical correlations are present, including long-range interactions \cite{ropke,biro,biro08}. Therefore, in the absence of a converging method to approach QCD at finite density one has often to resort to effective and phenomenological model investigations to obtain qualitative results.

In the last years there is a growing interest to high energy physics applications of
nonextensive statistics \cite{tsallis,GMTsallis,book2} and several authors have
outlined the possibility that experimental observations in relativistic heavy ion collisions can reflect nonextensive statistical behaviors
\cite{beck09,biro09,kodama09,wilk09,lavagno09,cley,chinellato}.

Nonextensive statistical mechanics introduced by Tsallis consists in a generalization of the common Boltzmann-Gibbs statistical mechanics and it is based upon the introduction of entropy \cite{book2}
\begin{eqnarray}
S_q[f]=\frac{1}{q-1}\, \left(1-\int[f({\bf x})]^q
\,d\Omega\right)\; ,\ \ \ \left(\int f({\bf x})\,d\Omega=1\right)
\, , \label{eq:GMTsallis}
\end{eqnarray}
where $f({\bf x})$ stands for a normalized probability distribution, ${\bf x}$ and $d\Omega$ denoting, respectively, a generic point and the volume element in the corresponding phase space (here and in the
following we set the Boltzmann and the Planck constant equal to unity). The real parameter $q$ determines the degree of non-additivity exhibited by the entropy form (\ref{eq:GMTsallis}) which reduces to the standard Boltzmann-Gibbs entropy in the limit $q\rightarrow 1$.
By means of maximizing the entropy $S_q$, under appropriate constraints, it is possible to obtain a probability distribution (or particle distribution) which generalized, in the classical limit, the Maxwell-Boltzmann distribution.
The nonextensive classical distribution can be seen as a superposition of Boltzmann one with different temperatures which have a mean value corresponding to the temperature appearing in the Tsallis distribution \cite{cley,wilkprl}.

The existence of nonextensive statistical effects incorporates in an effective way the presence of strong dynamical correlations, long-range interactions and microscopic memory effects and it affects significantly the finite temperature and density nuclear EOS \cite{miller2,epja2004,silva,wilknjl,epj2011,jpa2007}. In fact, by varying temperature and density, the EOS reflects in terms of the macroscopic thermodynamical variables the microscopic interactions of the different phases of nuclear matter.

In a previous work, we have studied the relevance of nonextensive statistical effects in the hadron to quark-gluon phase transition by means of a relativistic nuclear equation of state in a range of finite temperature and density for which the production of strange particle can be negligible and we have studied the deconfinement transition from nucleonic matter to up and down quark matter \cite{jpg2010}. By increasing the temperature of the system the producing of strangeness through associated production becomes much more important and we expect that nonextensive statistical effect considerably influence the strange particle densities in the EOS \cite{cent2012}.

The main goal of this paper is to study how nonextensive statistical effects influence, from a phenomenological point of view, the nuclear EOS in presence of strange matter and, as a consequence, the relative hadron to quark-gluon phase transition at finite temperature and baryon density reachable in high-energy heavy-ion collisions. At this scope we consider an effective relativistic mean-field model with the inclusion of the full octet of baryons, the $\Delta$-isobar degrees of freedom and the lightest pseudoscalar and vector mesons. These last particles are considered in the so-called one-body contribution, taking into account their effective chemical potentials depending on the self-consistent interaction between baryons \cite{muller,prc2010}.

During the phase transition, the Gibbs conditions for the phase equilibrium are applied by requiring the global conservation of three "charges": baryon number, electric charge (isospin) and strangeness number. A crucial feature of such conservation laws is that, during the mixed phase transition, the baryon, electric charge and strangeness densities can be different in the single hadron and quark phase, although the total net charge densities are kept fixed \cite{glenprd}. As a consequence, the pressure in the mixed phase is in general not constant and the nuclear incompressibility does not vanish at all \cite{Bonanno1}.

The nonextensive index $q$ is considered here as a free parameter, even if, in principle, it should depends on the physical condition generated in the reaction, on the fluctuation of the temperature and it should be related to microscopic quantities such as, for example, the mean interparticle interaction length, the screening length and the collision frequency into the parton plasma. Moreover, let us observe that when the entropic $q$ parameter is smaller than one, the high energy tail of the particle distribution is depleted; when $q$ is greater than one, the energy tail is enhanced. Hence the nonextensive statistics entails a sensible difference of the power-law particle distribution shape in the high energy region with respect to the standard statistics. In the present work we will focus our study for small deviations from the standard statistics and for values of $q>1$, because these values were obtained in several phenomenological studies in high energy heavy ion collisions (see, for example, Ref.s \cite{cley,chinellato,physica}). In this context, let us remember that, in the diffusional approximation, a value $q>1$ implies the presence of a superdiffusion among the constituent particles (the mean square displacement obeys to a power law behavior $\langle x^2\rangle\propto t^\alpha$, with $\alpha>1$) \cite{tsamem}.

The paper is organized as follows. In Sections \ref{hadron} and \ref{qgp}, we present, respectively, the main features of the effective nonextensive hadronic and quark-gluon EOS. In Section \ref{mp}, we report the Gibbs conditions applied for the hadron to quark-gluon phase transition and the consequent formation of a mixed phase. Finally, in Sections \ref{result} and \ref{conclusion}, we present the main results and summarize our conclusions.

\section{Nonextensive hadronic equation of state}
\label{hadron}

The relativistic mean-field model is widely successful used for describing the properties of finite nuclei as well as hot and dense nuclear matter \cite{walecka,boguta,c,glen}. In this framework, the Lagrangian density describing hadronic matter can be written as
\begin{equation}
{\mathcal L}={\mathcal L}_{\rm octet}+{\mathcal L}_\Delta+{\mathcal
L}_{\rm qpm} \, ,
\end{equation}
where ${\mathcal L}_{\rm octet}$ stands for the full octet of the
lightest baryons ($p$, $n$, $\Lambda$, $\Sigma^+$, $\Sigma^0$,
$\Sigma^-$, $\Xi^0$, $\Xi^-$) interacting with $\sigma$, $\omega$,
$\rho$ meson fields; ${\mathcal L}_\Delta$ corresponds to the degrees of freedom for the $\Delta$-isobars ($\Delta^{++}$,
$\Delta^{+}$, $\Delta^0$, $\Delta^-$) and ${\mathcal L}_{\rm qpm}$ is related to a gas of the lightest pseudoscalar and
vector mesons with an effective chemical potentials (see below for details).

The Lagrangian for the self-interacting octet of baryons can
be written as \cite{glen}
%
\begin{eqnarray}\label{lagrangian}
{\mathcal L}_{\rm octet} &=&
\sum_k\overline{\psi}_k\,[i\,\gamma_{\mu}\,\partial^{\mu}-(M_k-
g_{\sigma k}\,\sigma) -g_{\omega
k}\,\gamma_\mu\,\omega^{\mu}-g_{\rho k}\,\gamma_{\mu}\,\vec{t}
\cdot \vec{\rho}^{\;\mu}]\,\psi_k \nonumber\\
&&+\frac{1}{2}(\partial_{\mu}\sigma\partial^{\mu}\sigma-m_{\sigma}^2\sigma^2)
-\frac{1}{3}a\,(g_{\sigma
N}\,\sigma)^{3}-\frac{1}{4}\,b\,(g_{\sigma N}\,\sigma^{4})\nonumber\\
&& +\frac{1}{2}\,m^2_{\omega}\,\omega_{\mu}\omega^{\mu}
+\frac{1}{4}\,c\,(g_{\omega N}^2\,\omega_\mu\omega^\mu)^2
+\frac{1}{2}\,m^2_{\rho}\,\vec{\rho}_{\mu}\cdot\vec{\rho}^{\;\mu}\nonumber\\
&&-\frac{1}{4}F_{\mu\nu}F^{\mu\nu}
-\frac{1}{4}\vec{G}_{\mu\nu}\vec{G}^{\mu\nu}\,,
\end{eqnarray}
where the sum runs over the full octet of baryons, $M_k$ is the
vacuum baryon mass of index $k$, the quantity $\vec{t}$ denotes
the isospin operator that acts on the baryon and the field
strength tensors for the vector mesons are given by the usual
expressions
$F_{\mu\nu}\equiv\partial_{\mu}\omega_{\nu}-\partial_{\nu}\omega_{\mu}$,
$\vec{G}_{\mu\nu}\equiv\partial_{\mu}\vec{\rho}_{\nu}-\partial_{\nu}\vec{\rho}_{\mu}$.

In regime of finite values of temperature and density, a state of high density resonance matter may be formed and the $\Delta(1232)$-isobar degrees of freedom are expected to play a central role \cite{hofmann,zabrodin,bass,fachini}. In particular, the formation of resonances matter contributes essentially to baryon stopping, hadronic flow effects and enhanced strangeness \cite{mattiello}. The Lagrangian density concerning the $\Delta$-isobars can be expressed as \cite{greiner87,greiner97}
\begin{eqnarray}
{\mathcal L}_\Delta=\overline{\psi}_{\Delta\,\nu}\, [i\gamma_\mu
\partial^\mu -(M_\Delta-g_{\sigma\Delta}
\sigma)-g_{\omega\Delta}\gamma_\mu\omega^\mu
 ]\psi_{\Delta}^\nu \,
,
\end{eqnarray}
where $\psi_\Delta^\nu$ is the Rarita-Schwinger spinor  for the $\Delta$-baryon. Due to the uncertainty on the meson-$\Delta$ coupling constants, we limit ourselves to consider
only the coupling with the $\sigma$ and $\omega$ meson fields ($g_{\sigma\Delta}$ and $g_{\omega\Delta}$), more of which are explored in the literature \cite{greiner97,kosov,jin}.

The field equations in a mean field approximation are
\begin{eqnarray}
&&(i\gamma_{\mu}\partial^{\mu}-M^*_i-
g_{\omega i}\gamma^{0}\omega-g_{\rho i}\gamma^{0}{t_{3 i}}\rho)\psi_k=0\,, \\
&&(i\gamma_{\mu}\partial^{\mu}-M^*_{\Delta}-
g_{\omega \Delta}\gamma^{0}\omega)\psi_{\Delta}=0\,, \\
&&m_{\sigma}^2\sigma+ a g^3_{\sigma N}{{\sigma}^2}+ b g^4_{\sigma N}{{\sigma}^3}=\sum_{i}g_{\sigma i}\rho^s_i\,, \\
&&m^2_{\omega}\omega +cg^4_{\omega N}\omega^3=\sum_{i}g_{\omega i}\rho^B_i\,, \\
&&m^2_{\rho}\rho=\sum_{i}g_{\rho i}\tau_{3 i}\rho^B_i\, ,
\end{eqnarray}\label{eq:MFT}
where $\sigma=\langle\sigma\rangle$, $\omega=\langle\omega^0\rangle$, $\rho=\langle\rho^0_3\rangle$ and $\delta=\langle\delta_3\rangle$, are the nonvanishing expectation values of meson fields and the effective mass of the $i$-th baryon is defined as
\begin{eqnarray}
M^*_i= M_i- g_{\sigma i}\sigma \, .  \label{eq:Meff}
\end{eqnarray}
The $\rho^B_i$ and $\rho^S_i$ are the baryon
density and the baryon scalar density, respectively. They are
given by \cite{pla2002}
\begin{eqnarray}
&&\rho^{B}_i= \gamma_i \int\frac{{\rm
d}^3k}{(2\pi)^3}[n_i(k)-\overline{n}_i(k)]\,, \label{eq:rhob} \\
&&\rho^S_i= \gamma_i \int\frac{{\rm
d}^3k}{(2\pi)^3}\,\frac{M_i^*}{E_i^*}\,
[n_i^q(k)+\overline{n}_i^{\,q}(k)] \, .  \label{eq:rhos}
\end{eqnarray}
where $\gamma_i=2J_i+1$ is the degeneracy spin factor ($\gamma_{octet}=2$ for the baryon octet
and $\gamma_{\Delta}=4$) and $n_i(k)$ and $\overline{n}_i(k)$ are the $q$-deformed
fermion particle and antiparticle distributions.
Following Ref. \cite{wilknjl}, we can write, for $\beta(E_i^*(k)-\mu_i^*)>0$,
\begin{eqnarray}
n_i(k)=\frac{1} { [1+(q-1)\,\beta(E_i^*(k)-\mu_i^*)
]^{1/(q-1)} + 1} \label{eq:distribuz} \, ,
\end{eqnarray}
and, for $\beta(E_i^*(k)-\mu_i^*)\le 0$,
\begin{eqnarray}
n_i(k)=\frac{1} { [1+(1-q)\,\beta(E_i^*(k)-\mu_i^*)
]^{1/(1-q)} + 1} \label{eq:distribuz} \, .
\end{eqnarray}
The corresponding antiparticle distribution $\overline{n}_i$ can be obtained with the substitution $\mu_i^*\rightarrow -\mu_i^*$.
The baryon effective energy is defined as ${E_i}^*(k)=\sqrt{k^2+{{M_i}^*}^2}$ and the effective particle chemical
potentials $\mu_i^*$  are given in terms of the meson fields as follows
\begin{eqnarray}
\mu_i^*={\mu_i}-g_{\omega i}\omega - g_{\rho i} t_{3i}\rho \, ,
\label{mueff}
\end{eqnarray}
where $\mu_i$ are the thermodynamical chemical potentials
$\mu_i=\partial\epsilon/\partial\rho_i$ and $t_{3 i}$ is the third component of the isospin of $i$-th
baryon.

Because of we are going to describe a finite temperature and density
nuclear matter with respect to strong interaction, we have to
require the conservation of three "charges": baryon
number ($B$), electric charge ($C$) and strangeness number ($S$).
For this reason, the system is
described by three independent chemical potentials: $\mu_B$,
$\mu_C$ and $\mu_S$, respectively, the baryon, the electric charge
and the strangeness chemical potentials. Therefore,
the chemical potential of particle of index $i$ can be written as
\begin{equation}
\mu_i=b_i\, \mu_B+c_i\,\mu_C+s_i\,\mu_S \, . \label{mu}
\end{equation}
where $b_i$, $c_i$ and $s_i$ are, respectively, the baryon, the
electric charge and the strangeness quantum numbers of the $i$-th
hadronic species.

The meson fields are obtained as a solution of the field
equations in the mean field approximation and the related meson-nucleon
couplings constant ($g_{\sigma N}$, $g_{\omega N}$ and $g_{\rho N}$)
will be fixed to the parameters set marked as GM3 of Ref. \cite{glen}.
The implementation of the model with the inclusion of hyperons,
require the determination of the corresponding meson-hyperon
coupling constant, that have to be fitted to the hypernuclear proprieties.
This can be done fixing the scalar $\sigma$-meson hyperon coupling
constants to the potential depth of the corresponding hyperon in
saturated nuclear matter ($U_{\Lambda}^N=-28$ MeV, $U_{\Sigma}^N= 30$ MeV,
$U_{\Xi}^N=-18$ MeV) \cite{schaffner}.
The obtained ratios $x_{\sigma Y}=g_{\sigma Y}/g_{\sigma N}$ for the GM3 parameter set are:
$x_{\sigma \Lambda}=0.606$, $x_{\sigma \Sigma}=0.328$ and $x_{\sigma \Xi}=0.322$.
Furthermore, following
Ref. \cite{schaff4}, the SU(6) simple quark model can be used to obtain the relations
\begin{eqnarray}
\frac{1}{3}g_{\omega N}=\frac{1}{2}g_{\omega
\Lambda}=\frac{1}{2}g_{\omega \Sigma}=g_{\omega\Xi} \, , \nonumber\\
g_{\rho N}=\frac{1}{2}g_{\rho\Sigma}=g_{\rho\Xi}\,, \ \ \ g_{\rho
\Lambda}=0 \, .
\end{eqnarray}

Concerning the formation of $\Delta$-isobar matter at finite temperature and density, it has been predicted that a phase transition from nucleonic matter to $\Delta$-excited nuclear matter can take place and the occurrence of this transition sensibly depends on the $\Delta$-meson fields coupling constants \cite{greiner97,kosov}.
Whether metastable $\Delta$-excited nuclear matter exists or not is still a controversial issue because little is actually known about the $\Delta$-coupling constants with the scalar and vector mesons at finite temperature and density. On the other hand, QCD finite-density sum rule results predict a comparable or larger net attraction for a $\Delta$-isobar than for a nucleon in the nuclear medium \cite{jin}.
Because in this investigation we are going to focus on the nonextensive effects in hot and dense nuclear matter, we will not explore the existence of metastable $\Delta$-matter and the scalar and vector coupling ratio will be fixed equal to nucleon one (${g_{\sigma\Delta}}={g_{\sigma N}}$ and ${g_{\omega\Delta}}={g_{\omega N}}$) (see, for example, Ref. \cite{prc2010} for more details about different values of $\Delta$-coupling and the formation of $\Delta$-metastable matter for different EOSs).

The baryon pressure $P_B=-\Omega_B/V$ and the energy density $\epsilon$ can be obtained from the baryon grand potential $\Omega_B$ in the standard way. More explicitly, we have
\begin{eqnarray}
P_B&=&\frac{1}{3}\sum_i \,\gamma_i\,\int \frac{{\rm d}^3k}{(2\pi)^3}
\;\frac{k^2}{E_{i}^\star(k)}\; [n^q_i(k)+\overline{n}^{\,q}_i(k)]
-\frac{1}{2}\,m_\sigma^2\,\sigma^2 - U(\sigma)+
\frac{1}{2}\,m_\omega^2\,\omega^2  \nonumber\\
&+&\frac{1}{4}\,c\,(g_{\omega N}\,\omega)^4
+\frac{1}{2}\,m_{\rho}^2\,\rho^2\, \label{eq:eos},\\
\epsilon_B&=&\sum_i \,\gamma_i\,\int \frac{{\rm
d}^3k}{(2\pi)^3}\;E_{i}^\star(k)\; [n^q_i(k)+\overline{n}^{\,q}_i(k)]
+\frac{1}{2}\,m_\sigma^2\,\sigma^2
+U(\sigma)+\frac{1}{2}\,m_\omega^2\,\omega^2 \nonumber\\
&+&\frac{3}{4}\,c\,(g_{\omega N}\,\omega)^4 +\frac{1}{2}\,m_{\rho}^2
\,\rho^2 \, .
\label{eq:eos2}
\end{eqnarray}

The contribution of the lightest pseudoscalar ($\pi$, $K$, $\overline{K}$, $\eta$,
$\eta'$) and vector mesons ($\rho$, $\omega$, $K^*$, $\overline{K}^*$, $\phi$) to the thermodynamics
potential and to the others thermodynamics quantities, in the regime of temperatures and baryon densities explored in this work, is not negligible.
On the other hand, the contribution of the $\pi$ mesons (and other
pseudoscalar and pseudovector fields) vanishes at the mean-field
level. Following Ref.s \cite{muller,prc2010}, we can evaluate the
pressure $P_M$, the energy density $\epsilon_M$ and particle density $\rho_M$ as that of a
quasi-particle gas in the so-called one-body contribution, taking into account their effective chemical potentials depending on the self-consistent interaction between baryons. In this approximation we can write
\begin{eqnarray}
&&P_M= \frac{1}{3}\sum_j \,\gamma_j\,\int \frac{{\rm
d}^3k}{(2\pi)^3} \;\frac{k^2}{E_{j}(k)}\; [g^q_j(k)+\overline{g}^{\,q}_j(k)]\, ,
\label{pmeson}\\
&&\epsilon_M=\sum_j \,\gamma_j\,\int \frac{{\rm
d}^3k}{(2\pi)^3}\;E_{j}(k)\; [g^q_j(k)+\overline{g}^{\,q}_j(k)] \, ,
\label{emeson}\\
&&\rho_j^M=\gamma_j \int\frac{{\rm
d}^3k}{(2\pi)^3}\;[g_j(k)-\overline{g}_j(k)]  \, .
\label{rhomeson}
\end{eqnarray}
where $\gamma_j=2J_j+1$ is the degeneracy spin factor and the functions $g_j(k)$ and $\overline{g}_j(k)$ are the $q$-deformed particle and anti-particle distribution functions of the $j$-th meson. Again, for $\beta(E_j(k)-\mu_j^*)>0$ we have
\begin{eqnarray}
g_j(k)=\frac{1} { [1+(q-1)\,\beta(E_j(k)-\mu_j^*)
]^{1/(q-1)} + 1} \label{eq:distribuzmes} \, ,
\end{eqnarray}
where $E_j(k)=\sqrt{k^2+m_j^{2}}$.
Moreover, the boson integrals are subjected to the constraint $\vert\mu_j^*\vert\le m_j$, otherwise Bose condensation becomes possible.

The values of the effective meson chemical potentials $\mu_j^*$ are obtained from the "bare" ones $\mu_j$, given in Eq.(\ref{mu}), and subsequently expressed in terms of the corresponding effective baryon chemical potentials, respecting the strong interaction \cite{muller,prc2010}. More explicitly, we have from Eq.(\ref{mu}) that $\mu_{\pi^+}=\mu_{\rho^+}=\mu_C\equiv\mu_p-\mu_n$ and the
corresponding effective chemical potential can be written as
\begin{eqnarray}
\mu_{\pi^+(\rho^+)}^*\equiv\mu_p^*-\mu_n^* = \mu_p-\mu_n-g_{\rho N}\,\rho \, , \label{mueff_m1}
\end{eqnarray}
where the last equivalence follows from Eq.(\ref{mueff}).

Analogously, by setting $x_{\omega \Lambda}=g_{\omega\Lambda}/g_{\omega N}$, we have
\begin{eqnarray}
\mu_{K^+(K^{*+})}^* &\equiv\mu_p^*-\mu_{\Lambda(\Sigma^0)}^* = \mu_p-\mu_{\Lambda}- (1-x_{\omega \Lambda})g_{\omega N
}\omega-\frac{1}{2}\,g_{\rho N}\rho \, ,\label{mueff_m2}\\
\mu_{K^0(K^{*0})}^* &\equiv\mu_n^*-\mu_{\Lambda(\Sigma^0)}^* =\mu_n-\mu_{\Lambda}- (1-x_{\omega \Lambda})g_{\omega N
}\omega+ \frac{1}{2}\,g_{\rho N}\rho \, , \label{mueff_m3}
\end{eqnarray}
while the others strangeless neutral mesons have a vanishing chemical potential. Thus, the effective meson chemical potentials are coupled with the meson fields related to the interaction between baryons.
This assumption represents a crucial feature in the EOS at finite density and temperature and can be seen somehow in analogy with the hadron resonance gas within the excluded-volume approximation. There the hadronic system is still regarded as an ideal gas but in the volume reduced by the volume occupied by constituents (usually assumed as a phenomenological model parameter), here we have a (quasifree) meson gas but with an effective chemical potential that contains the self-consistent interaction of the meson fields.

Finally, the total pressure and energy density can be written as the
sum of the baryon and meson contributions: $P_{\rm tot}=P_B+P_M$, $\epsilon_{\rm tot}=\epsilon_B+\epsilon_M$.

\section{Nonextensive QGP equation of state}\label{qgp}

Concerning the nonextensive quark-gluon EOS, we use a simple effective MIT bag model to describe the quark-gluon phase in which all the non-perturbative effects are simulated by
the bag constant $B$ which represents the pressure of the vacuum.
It is well known that, using the simplest version of the MIT bag
model, at moderate temperatures the deconfinement transition takes
place at very large densities if the bag pressure $B$ is fixed to
reproduce the critical temperature computed in lattice QCD. On the
other hand there are strong theoretical indications that at large densities (and not too large temperatures)
diquark condensates can form, whose effect can be approximately
taken into account by reducing the value of the effective bag
constant \cite{Drago1}. A phenomenological approach can therefore be based
on a bag constant depending on the baryon chemical potential \cite{Bonanno1}. In this study we consider a parametrization of the form
\begin{eqnarray}
B_{\rm eff}= (B_0 - B_\infty)/(1 + \exp[(\mu_B - \mu_0)/a])   + B_\infty \, ,
\end{eqnarray}
where we have fixed $B_0^{1/4}=254$ MeV (bag constant at vanishing $\mu_B$), $B_\infty^{1/4}=160$ MeV (bag constant at very large $\mu_B$), $\mu_0=600$ MeV and $a=320$ MeV.
The above values have been obtained by requiring that, at vanishing chemical potential, the critical temperature is about  170 MeV for $q=1$, as suggested by lattice calculations \cite{karsch}, while the other constraint is the requirement that the mixed phase starts forming at a density slightly exceeding $3\,\rho_0$ for a temperature of the order of 100 MeV (as suggested by hydrodynamical calculations \cite{toneev}).

Following this line, the
pressure, energy density and baryon number density for a
relativistic Fermi gas of quarks can be written, respectively, as \cite{pla2002}
\begin{eqnarray}
&P& =\frac{\gamma_f}{3} \sum_{f} \int^\infty_0 \frac{{\rm
d}^3k}{(2\pi)^3} \,\frac{k^2}{e_f}\, [n^q_f(k)+\overline{n}^{\,q}_f(k)]
-B_{\rm eff}\,, \label{bag-pressure}\\
&\epsilon& =\gamma_f \sum_{f}  \int^\infty_0 \frac{{\rm
d}^3k}{(2\pi)^3} \,e_f\, [n^q_f(k)+\overline{n}^{\,q}_f(k)]
\label{bag-energy}
+B_{\rm eff}\,, \\
&\rho& =\frac{\gamma_f}{3} \sum_{f} \int^\infty_0 \frac{{\rm
d}^3k}{(2\pi)^3}  \,[n_f(k)-\overline{n}_f(k)]\, ,
\label{bag-density}
\end{eqnarray}
where the quark degeneracy factor for each flavor ($f=u,d,s$) is
$\gamma_f=6$ and $n_f(k)$, $\overline{n}_f(k)$ are the
$q$-deformed particle and antiparticle quark distributions. More explicitly,
\begin{eqnarray}
n_f(k)=\frac{1} { [1+(q-1) \,\beta(e_f(k)-\mu_f)
]^{1/(q-1)} + 1} \, ,
\end{eqnarray}
for $\beta(e_f(k)-\mu_f)>0$, where $e_f(k)=\sqrt{k^2+m_f^2}$.

Light quarks $u$ and $d$ are considered as massless particles, while for strange quarks we consider a finite mass of $m_s=150$ MeV.
Similar expressions for the pressure and the energy density can be written for gluons treating them as a massless $q$-deformed Bose gas with zero chemical potential. Therefore, the nonextensive pressure $P_g$ and energy density $\epsilon_g$ for gluons can be written as
\begin{eqnarray}
&P_g& =\frac{\gamma_g}{3} \int^\infty_0 \frac{{\rm
d}^3k}{(2\pi)^3}
\,\frac{k}{[1+(q-1)\,k/T]^{q/(q-1)} - 1}\,, \label{gluon-press}\\
&\epsilon_g& =3\, P_g \, , \label{gluon-energy}
\end{eqnarray}
with the gluon degeneracy factor $\gamma_g=16$. In the limit
$q\rightarrow 1$, one recovers the usual analytical expression:
$P_g=(8\pi^2/45)\,T^4$.

\section{Mixed phase of hadron and quark-gluon matter}\label{mp}

As briefly mentioned in the Introduction, by increasing the baryon chemical potential, the phase diagram is characterized by a rapid crossover of thermodynamic observable around a quasi-critical temperature with a CEP, beyond which the system shows a first-order phase transition from confined to deconfined matter.

The main goal of this study is to show the relevance of nonextensive statistical effects in a finite range of temperature and baryon density relevant for compressed baryonic matter experiments rather than the ultrarelativistic regime at vanishing baryon chemical potential. Thus, by considering the deconfinement transition at finite density, a mixed phase can be formed, which is typically described using the two separate equations of state: one for the hadronic phase and one for the quark-gluon phase.

In order to study the phase transition from hadronic matter to QGP, we apply the Gibbs conditions to systems with more than one conserved charge, by requiring the global conservation of the baryon number, electric charge and zero net strangeness. The structure of the mixed phase is therefore obtained by imposing the following conditions
\begin{eqnarray}
&&\mu_B^{(H)} = \mu_B^{(Q)}\,, \ \ \mu_C^{(H)} = \mu_C^{(Q)}\,, \ \ \mu_S^{(H)} = \mu_S^{(Q)} \,,\nonumber\\
&&P^{(H)}(T,\mu_B,\mu_C,\mu_S)=P^{(Q)}(T,\mu_B,\mu_C,\mu_S)\, ,
\nonumber \\
&&\rho_B=(1-\chi)\,\rho_B^H(T,\mu_B,\mu_C,\mu_S)+\chi\, \rho_B^Q(T,\mu_B,\mu_C,\mu_S)\,, \\
&&\rho_C=(1-\chi)\,\rho_C^H(T,\mu_B,\mu_C,\mu_S)+\chi\, \rho_C^Q(T,\mu_B,\mu_C,\mu_S) \, ,\nonumber \\
&&\rho_S=(1-\chi)\,\rho_S^H(T,\mu_B,\mu_C,\mu_S)+\chi\, \rho_S^Q(T,\mu_B,\mu_C,\mu_S) \, ,\nonumber
\end{eqnarray}
where $\rho_B^{H(Q)}$, $\rho_C^{H(Q)}$ and $\rho_S^{H(Q)}$ are, respectively,  the baryon, electric charge and strange densities in the hadronic (H) and in the quark (Q) phase and $\chi$ is the volume fraction of quark-gluon matter in the mixed phase. At fixed $T$ and $\mu_B$, the electric charge $\mu_C$ and strangeness $\mu_S$ chemical potentials are obtained by fixing the net electric charge $Z/A$ (for example, $Z/A=0.4$ for lead-lead heavy ion collisions) and the net  zero strangeness condition.

As previously discussed, an important aspect it that the presence of more than one conserved charge implies a global and not local charge conservation, therefore the charge densities $\rho_B$, $\rho_C$ and $\rho_S$ are fixed only as long as the system remains in one of the two pure phases. In the mixed phase, the charge concentration in each of the regions of one phase or the other may be different. As we will see, this feature plays a crucial role on the strangeness production during the mixed phase of hadron and quark-gluon matter.

\section{Results and discussion}\label{result}

\begin{figure}
\begin{center}
\resizebox{0.6\textwidth}{!}{%
\includegraphics{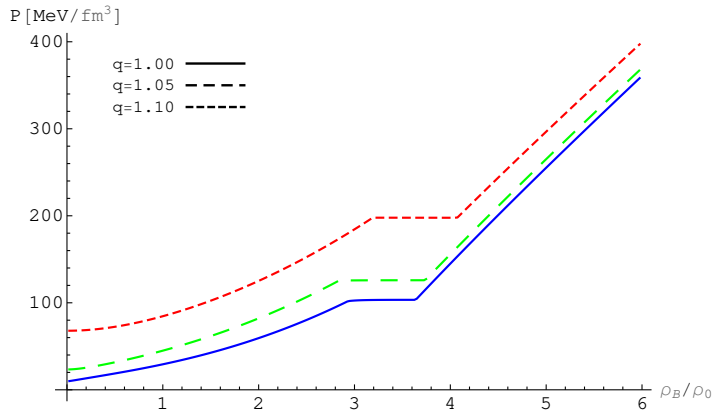}
} \caption{Pressure as a function of baryon density (in units of the nuclear saturation density $\rho_0$) at $T=120$ MeV and for different values of nonextensive parameter $q$.} \label{fig:Pressrhob}
\end{center}
\end{figure}

Let us start our numerical investigation by reporting in figure \ref{fig:Pressrhob} the variation of the pressure as a function of baryon density (in units of nuclear saturation density $\rho_0=0.153$ fm$^{-3}$), at $T = 120$ MeV and for different values of $q$. Here and in the following we fix the value $Z/A = 0.4$. In presence of nonextensive statistical effects the pressure results to be considerably increased even for small deviations from the Boltzmann-Gibbs statistics. It is interesting to observe that the pressure presents a strong softening in the mixed phase even if the Gibbs conditions on the phase transition are applied. At this temperature, such a behavior results to be more pronounced by increasing the value of the nonextensive entropic parameter $q$ with a larger range of baryon density involved in the mixed phase region. This matter of fact, already present in absence of strange particle degrees of freedom \cite{jpg2010}, implies an abrupt variation in the incompressibility and results much more evident here due to the additional strangeness conservation constraint. In this context let us observe that indirect indications of a remarkable softening of the EOS at energies reached at AGS have been already outlined \cite{ivanov}.
\begin{figure}
\begin{center}
\resizebox{0.6\textwidth}{!}{%
\includegraphics{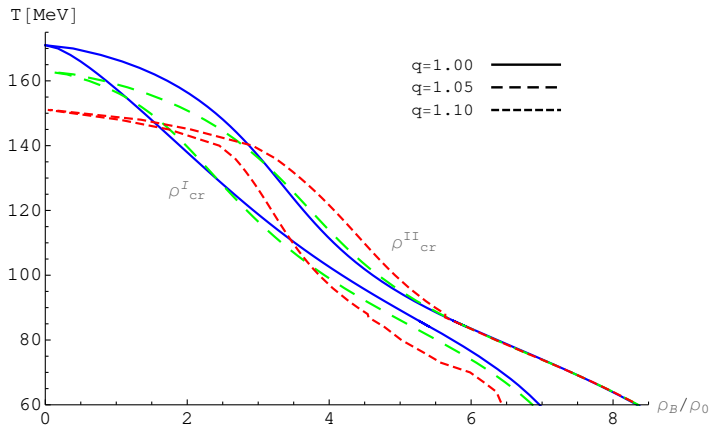}
} \caption{Phase diagram in the $\rho_B - T$ plane for different values of $q$.} \label{fig:Trhoc}
\end{center}
\end{figure}

In figure \ref{fig:Trhoc}, we report the phase diagram in the $\rho_B - T$ plane
for different values of the nonextensive parameter $q$. The curves labeled with
$\rho^{I}_{cr}$ and $\rho^{II}_{cr}$ denote, respectively, the beginning and the
end of the mixed phase. In presence of nonextensive statistical effects the phase diagram results significantly  modified and a remarkable lowering of the critical maximum temperature at vanishing baryon density $\rho_B$ is present. This result is in according to previous investigations where, by fitting the experimental observable at $q>1$, the temperature (or slope) parameter $T$ is usually less that the one obtained in the standard Boltzmann-Gibbs statistics ($q=1$) \cite{cley,physica,wilk2}.

\begin{figure}
\begin{center}
\resizebox{1\textwidth}{!}{%
\includegraphics{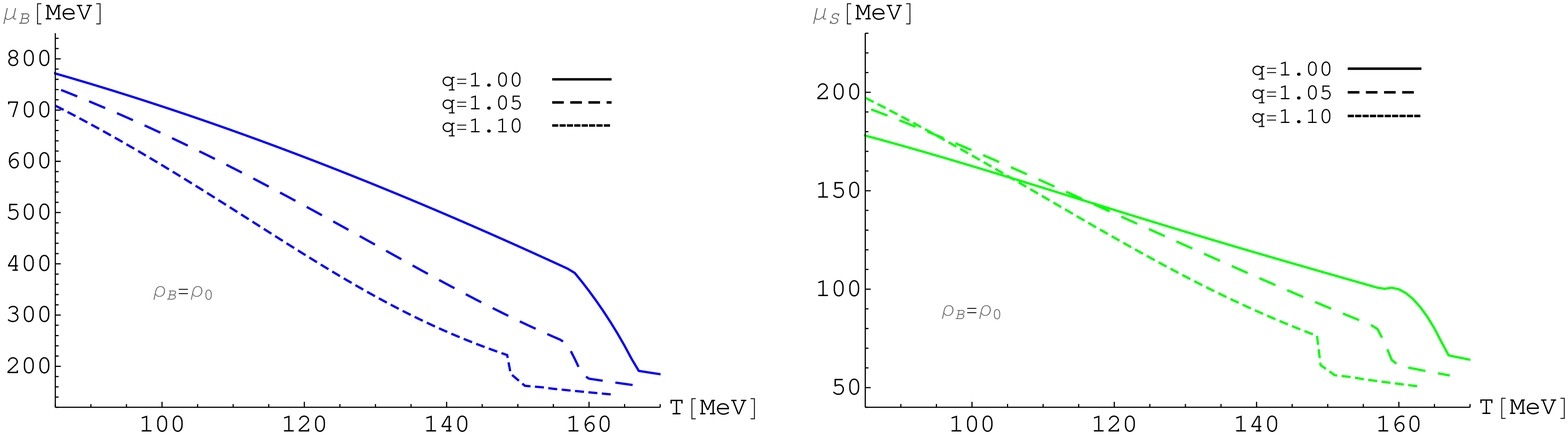}
} \caption{Variations of the baryon chemical potential $\mu_B$ (left panel) and the strangeness chemical potential $\mu_S$ (right panel) as a function of temperature at $\rho_B=\rho_0$ and for different values of $q$.} \label{fig:muitemp}
\end{center}
\end{figure}

In figure \ref{fig:muitemp}, we show the variation of the baryon chemical potential $\mu_B$ (left panel) and the strangeness chemical potential $\mu_S$ (right panel) as a function of temperature at fixed $\rho_B=\rho_0$ and for different values of $q$.
As expected, for a multicomposed strange hadronic matter, $\mu_S$ is positive and decreased with $T$ at a fixed $\rho_B$. It is interesting to observe the different behavior of $\mu_S$ at lower and higher temperatures in the presence ($q\ne 1$) and in absence ($q=1$) of nonextensive statistical effects.

\begin{figure}
\begin{center}
\resizebox{1\textwidth}{!}{%
\includegraphics{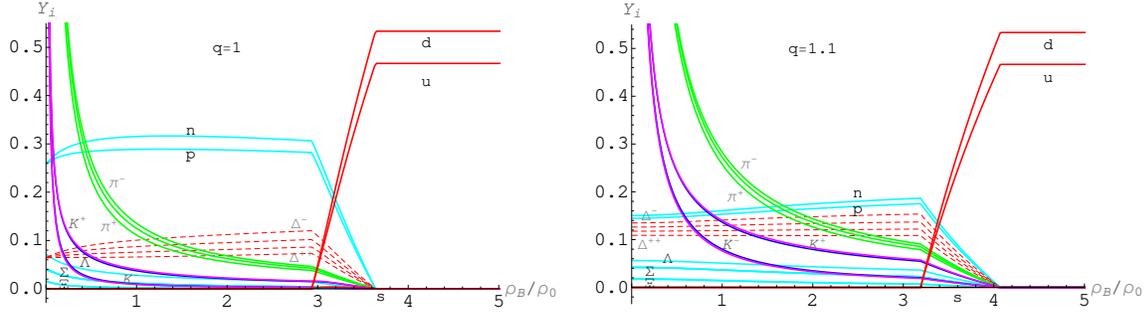}
} \caption{Particles concentration as a function of baryon density at $T = 120$ MeV and for $q = 1$ (left panel) and $q=1.1$ (right panel).} \label{fig:Yi}
\end{center}
\end{figure}

Let us now explore in more detail the particle concentrations for different values of the nonextensive parameter $q$. In figure \ref{fig:Yi}, we report the most relevant net particle ratios ($Y_i=\rho_i/\rho_B$) as a function of baryon density at $T$=120 MeV. In presence of nonextensive statistical effects (right panel), the net particle concentrations are sensibly modified even for small deviations from the standard statistics. In particular, we observe in the hadronic phase a strong reduction of the neutron and proton fractions and a considerable increase in the hyperon and in the meson concentrations, also at moderate baryon densities.

\begin{figure}[htb]
\begin{center}
\resizebox{1\textwidth}{!}{%
\includegraphics{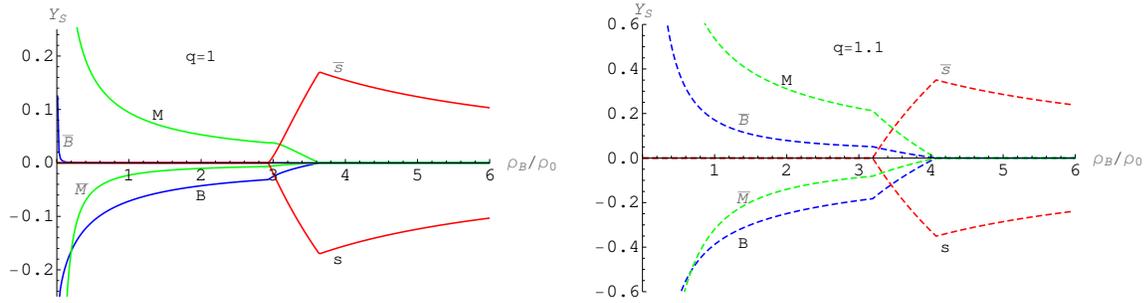}
} \caption{Strangeness fractions of baryons ($B$), mesons ($M$), strange quarks ($s$) and their antiparticles as a function of baryon density in the pure hadronic phase, mixed phase and quark phase at $T$=120 MeV for $q = 1$ (left panel) and $q=1.1$ (right panel).} \label{fig:Ys}
\end{center}
\end{figure}

To better understand the relevance of the nonextensive statistical effects in presence of strange
matter, we show in figure \ref{fig:Ys} the strangeness fraction $Y_S$ for baryons ($B$), mesons ($M$), strange quarks ($s$) and their antiparticles as a function of baryon density at a fixed temperature of $T$=120 MeV. For $q=1$ (left panel), in the hadronic phase ($\chi=0$), the total strangeness is carried almost completely by mesons (kaons, mainly $K^+$ and $K^0$) and baryons (hyperons), although at low baryon density anti-mesons $\overline{M}$ (anti-kaons, mainly $K^-$ and $\overline{K}^0$) bring a non-negligible fraction of strangeness. At the beginning of the mixed phase ($\rho_B\approx 3\,\rho_0$) the onset of $s$ and $\overline{s}$ quarks rapidly
dominates the strangeness ratio and the contribution of the other particle fractions becomes gradually less relevant.

In presence of nonextensive statistical effects (right panel) the situation is quite different. Here we observe a strong enhancement in all strange-particle concentrations and, in particular, the contribution of anti-baryons $\overline{B}$ (anti-hyperons) becomes relevant and comparable to anti-mesons and baryons.
\begin{figure}[htb]
\begin{center}
\resizebox{1\textwidth}{!}{%
\includegraphics{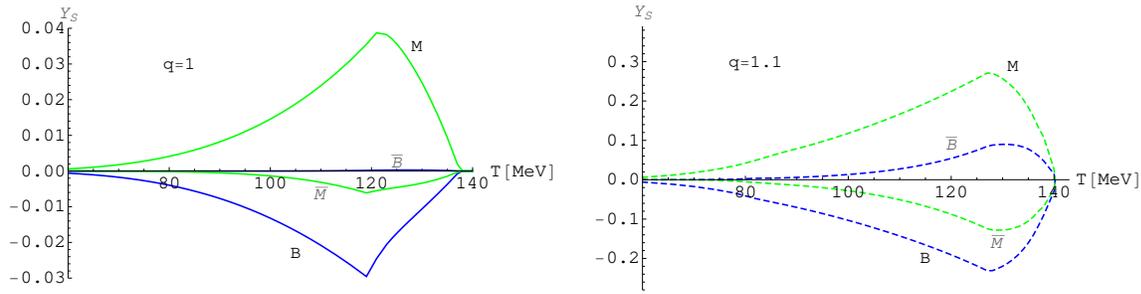}
} \caption{Strangeness fractions as a function of temperature in the pure
hadronic and mixed phase at $\rho_B=3\,\rho_0$ for $q = 1$ (left panel) and $q=1.1$ (right panel).} \label{fig:Ys_temp3rho}
\end{center}
\end{figure}

This behavior is also evident in figure \ref{fig:Ys_temp3rho}, where the strangeness fractions in the hadron and in the mixed phase as a function of the temperature at $\rho=3\,\rho_0$ are shown (quark phase is not reported in this figure).
As expected, by increasing the temperature all the strange particle densities are sensibly enhanced. Furthermore, in both panels we observe a peak at the end of the hadronic phase and a rapid decrease at the beginning of the mixed phase, due to the onset of the strange quarks.

The strong enhancement of the strange hadronic particles in presence of nonextensive statistical effects is a direct consequence of the normalized mean occupation function $n_i$ because, for $q>1$ and fixed baryon density (or $\mu_B$), it is enhanced at high values of its argument and depressed at low values. Being the argument of the mean occupation function $x_i=\beta (E_i^*-\mu_i^*)$, in the integration over momentum (energy), at lower $\mu_{B}$ (corresponding to lower values of $\mu_i^*$) the enhanced nonextensive high energy tail weighs much more that at higher $\mu_{B}$ where depressed low energy effects prevail and the mean occupation number results to be bigger for the standard Fermi-Dirac statistics. Concerning the antiparticle contribution, the argument of $\overline{n}_i$ is $\overline{x}_i=\beta (E_i^*+\mu_i^*)$ and the nonextensive enhancement at high energy tail is favored also at higher $\mu_B$. As a consequence the formation of antiparticles results to be enhanced in presence of nonextensive statistical effects. At the same time, higher temperatures reduce the value of the argument of $n_i$ and $\overline{n}_i$, reducing the effect of the nonextensive distribution function.

These properties change significantly the behavior of the $\sigma$ meson field and, therefore, the effective baryon mass which is related by the relation: $M^*_i=M_i-g_{\sigma i}\sigma$. In fact, for $q>1$, $M^*_i$ becomes smaller at low baryon density (and higher temperature) and bigger at finite densities $\rho_B\gtrsim 0.2 \, \rho_0$ (and lower temperature) \cite{jpg2010}. This effect appears to be much more significant as a percentage for lighter baryons, therefore, at finite baryon densities, the nucleons effective mass is enhanced compared to the standard case ($q=1$) with a percentage significantly greater than the hyperons effective masses. This matter of fact favors the formation of hyperons compared to  that of nucleons at a fixed baryon density and in presence of nonextensive statistical effects. Moreover, being the strangeness number globally conserved, an increase of hyperon particles implies, in the pure hadron phase, a corresponding increase of strange meson particles in order to satisfy the condition of zero net strangeness.

\begin{figure}[ht]
\begin{center}
\resizebox{1\textwidth}{!}{%
\includegraphics{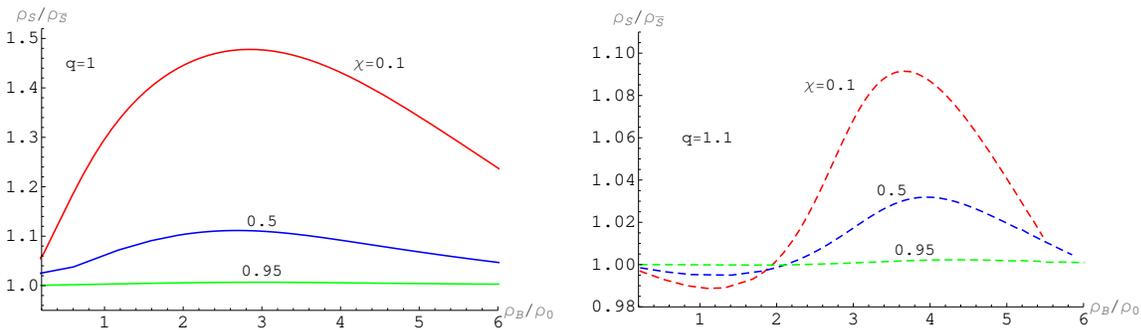}
} \caption{Ratio of strange to anti-strange quarks in the quark phase, as
a function of the baryon density at different values of the volume fraction of quark-gluon matter $\chi$ in the mixed phase for $q=1$ (left panel) and $q=1.1$ (right panel).}
\label{fig:quark_stranezza}
\end{center}
\end{figure}

At the scope of better focalize the role of nonextensive statistical effects in the strangeness production during the mixed phase, in figure \ref{fig:quark_stranezza}, we report the strange to anti-strange quark ratio $\rho_s/\rho_{\,\overline{s}}$ as a function of the baryon density, at different values of the volume fraction of quark matter $\chi$ in the quark-gluon phase along the phase transition, i.e., for a continuously varying temperature. In this context, let us point out again that, due to the Gibbs conditions, into the mixed phase, the net strangeness in each separate phase need not vanish although the total net strangeness is zero. This matter of fact is evident in figure \ref{fig:quark_stranezza}. In the left panel ($q=1$), especially at the beginning of the mixed phase ($\chi=0.1$), there is a remarkable excess of quark $s$ with respect to $\overline{s}$, with a maximum value around  $\rho_B\approx 3\,\rho_0$. This means that in absence of nonextensive statistical effects, during the phase transition, there is a large antistrangeness ($\overline{s}$) content in the hadron phase while quark-gluon plasma retains a large net strangeness ($s$) excess. This distillation mechanism, already known in literature, may result in "strangelet" formation, i.e. metastable droplets of strange-quark matter, which could imply a unique signature for quark-gluon plasma formation in relativistic heavy-ion collisions \cite{greiner0}.

\begin{figure}[ht]
\begin{center}
\resizebox{1\textwidth}{!}{%
\includegraphics{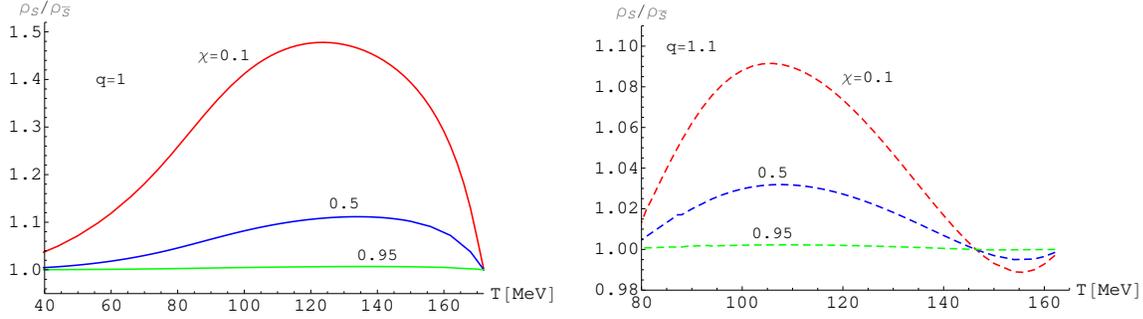}
} \caption{The same of figure \ref{fig:quark_stranezza} as a function of the temperature}
\label{fig:quark_stranezza2}
\end{center}
\end{figure}

In presence of the nonextensive statistical effects (right panel), there is a very different behavior with a ratio $\rho_s/\rho_{\,\overline{s}}$ slightly less than one at lower baryon density ($\rho_B< 2\,\rho_0$), whereas the other way round occurs at higher $\rho_B$ with a comparable behavior to the $q=1$ case but with an excess of $s$ quarks less pronounced. Similar behavior can be observed in figure \ref{fig:quark_stranezza2}, where the same ratio as a function of the temperature is also reported. Of course, at fixed $\chi$, low temperatures imply high densities and viceversa.

As previous discussed, this behavior is a consequence of the power-law behavior of the mean occupational distribution for $q>1$: the formation of antihyperons and kaons (mainly $K^+$, $K^0$) turns out to be disadvantaged at low baryon density and very high temperature and, viceversa, favored at high baryon density and low/intermediate temperature, compared to the standard case ($q=1$). Although particles with strangeness and antistrangeness content are strongly enhanced in presence of nonextensive statistics, the ratio strangeness/antistrangeness results depleted due to an enhancement of antiparticles  production at finite temperature.


\begin{figure}
\begin{center}
\resizebox{1\textwidth}{!}{%
\includegraphics{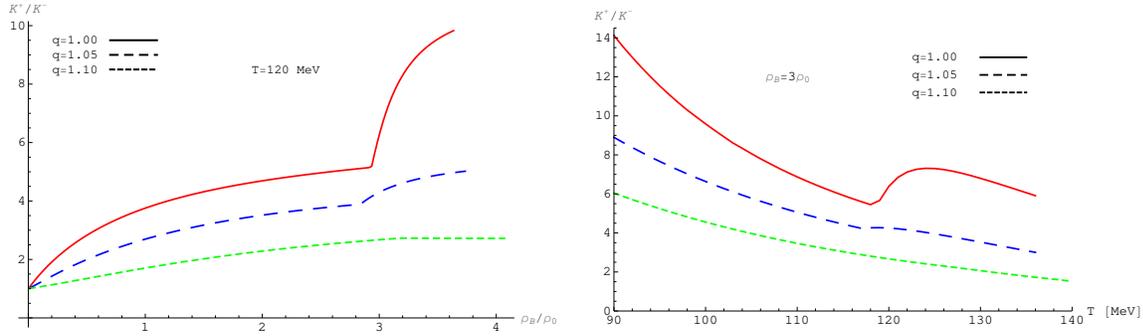}
} \caption{Kaon to antikaon ratio as a function of baryon density (left panel) and temperature (right panel) for different values of $q$ in the hadronic phase.} \label{fig:kk_temp}
\end{center}
\end{figure}

In agreement with the previous results, in figure \ref{fig:kk_temp}, we report the $K^+/K^-$ ratio in the hadronic phase as a function of baryon density at $T=120$ MeV (left panel) and as a function of temperature at $\rho_B=3\, \rho_0$ (right panel). For $q=1$ and fixed temperature, the $K^+/K^-$ ratio increases with continuity by increasing the baryon density until it reaches the first critical density around ($\rho_B\approx 3\,\rho_0$), after that a strong enhancement of the ratio occurs in the mixed phase region. This a consequence of the antistrangeness ($\overline{s}$) excess in the hadronic phase at the beginning of the mixed phase. For the same reason the $K^+/K^-$ ratio decreases by increasing the temperature until the beginning of the mixed phase with an abrupt increase of the ratio.
On the other hand, for $q>1$, the kaon to anti-kaons ratio is depleted with respect to the standard case ($q=1$), due to the increase of the antiparticles production. Furthermore, the strong variation of the ratio during the mixed phase transition at $q=1$ results to be very smoothed in presence of nonextensive statistical effects.

\begin{figure}
\begin{center}
\resizebox{1\textwidth}{!}{%
\includegraphics{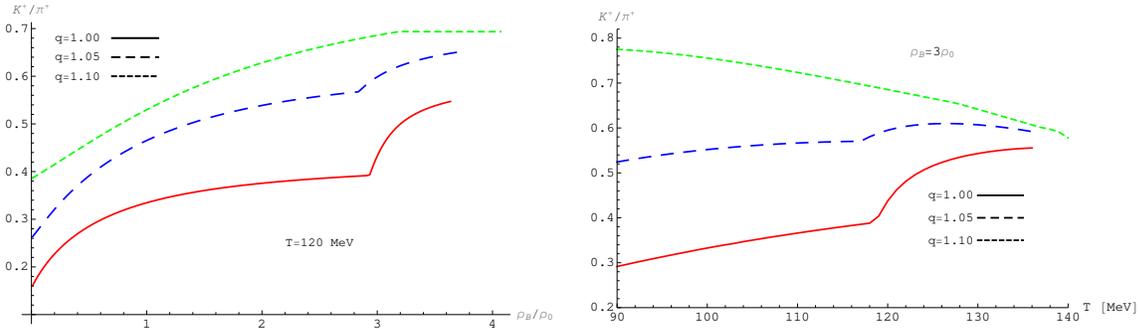}
} \caption{The same of figure \ref{fig:kk_temp} for the $K^+/\pi^+$ ratio.} \label{fig:kpi_temp}
\end{center}
\end{figure}

In presence of nonextensive statistical effects, the strong enhancement in density of strange particles compared to non-strange can be further observed in figure \ref{fig:kpi_temp} where the $K^+/\pi^+$ ratio is reported as a function of baryon density at $T=120$ MeV (left panel) and as a function of temperature at $\rho_B=3\, \rho_0$ (right panel). For $q=1.1$, there is a strong increase of the ratio, while it does not present discontinuities at the beginning of the mixed phase, unlike the case $q=1$ where a much more antistrangeness excess is present. Finally it is interesting to observe for $q=1.1$ a continuous decreasing trend of the ratio as a function of temperature at fixed baryon density.

\section{Conclusions}\label{conclusion}

We have studied an effective nuclear EOS in the framework of nonextensive statistical effects at finite temperature and baryon dentity. By requiring the Gibbs conditions on the global conservation of the baryon
number, electric charge fraction and zero net strangeness, we have studied the phase transition from hadronic matter to quark-gluon plasma and it has been shown that nonextensive statistical effects play a crucial role in the deconfinement phase transition also for small deviations from the standard Boltzmann-Gibbs statistics.
Our investigation is focalized in regime of finite temperature and baryon density relevant for future compressed baryonic matter experiments. As previously discussed, in regime of high temperature and very small baryon density (low $\mu_B)$, the phase transition may end in a second order critical endpoint with a smooth crossover and these features cannot be incorporated in the considered mean field approach.

We have shown that strange particles are abundantly produced even at moderate temperature and the anti-particles concentration is strongly increased in presence of nonextensive statistical effects. For $q>1$, the phase transition is characterized by an antistrangess content in the hadron phase while the QGP retains a net strangeness excess at large densities ($\rho_B\gtrsim 2\,\rho_0$) and intermediate temperatures ($T\approx 100\div 120$ MeV), while the other way round occurs at low densities and high temperature. This matter of fact is essentially due to the power law behavior of the mean occupational distribution function which weights differently low and high energy states at different baryon densities and temperatures.

The possibility of separating strange from antistrange matter in the hadron-QGP phase transition can lead to a very significant enrichment of strange quarks in the QGP at high baryon density and intermediate temperature. In fact, in the hadronic sector of the mixed phase, the $K^+$ and $K^0$ are enhanced and the hyperons are suppressed. Being mesons much lighter than nucleons and their resonances, they carry away entropy, energy and antistrangeness and, therefore, the prompt kaon emission cools and charges the system with finite net strangeness, leading to an even stronger enhancement of the $s$ quarks in the quark phase. This feature, which also depends strongly on the value of the strange particle densities, favors the formation of metastable or stable droplets of strange quark matter which would contain approximately the same amount of $u$, $d$ and $s$ quarks. The evidence of such a state of quark matter could be related to the existence of exotic hadron states, like a $H$-dibaryon state, a deeply bound 6-quark state predicted by Jaffe more than thirty years ago \cite{jaffe}. Although many experimental searches for the $H$-dibaryon were carried out and so far no convincing signal was found \cite{yoon_prc2007}, very recently evidence for a bound $H$-dibaryon was claimed based on lattice QCD calculations \cite{beane_prl2011,inoue_prl2011}.

In presence of nonextensive statistical effects, the separation of strange and antistrange quarks in the hadron-QGP mixed phase turns out to be less pronounced than in the standard case due to a more symmetric presence of particle and antiparticle at intermediate temperatures. On the other hand, for $q>1$, the strangeness fractions $Y_S$ (and, consequently, the densities of strange particles) result to be much greater compared to the $q=1$ case. This matter of fact could be crucial in the formation and survival of strange quark matter droplets in relativistic heavy ion collisions at high compressed baryonic matter.

\end{document}